\documentclass[
aps
,pra
,tightenlines
,twoside
    ,twocolumn
,superscriptaddress
,showpacs
]{revtex4-1}

\usepackage{graphicx}
\usepackage{color}
\usepackage{epsfig}
\usepackage{amstext}
\usepackage{amssymb}
\usepackage{bm}
\usepackage[english]{babel}
\usepackage[latin2]{inputenc}

\newcommand{\fer}[2]{c_{#1#2}}

\newcommand{\sigmab}{\bar{\sigma}}

\begin{document}
\author{Przemys{\l}aw R. Grzybowski and Ravindra W. Chhajlany}
\address{Faculty of Physics, Adam Mickiewicz University, Umultowska 85, 61-614 Pozna{\'n}, Poland}


\date{\today}
\title{Hubbard-I approach to the Mott transition}

\begin{abstract}
We expose the relevance of double occupancy conservation symmetry in
application of the Hubbard-I approach to strongly correlated electron
systems.  We propose the utility of a composite method, {\it viz.} the
Hubbard-I method in conjunction with strong coupling perturbation
expansion, for studying systems violating the afore--mentioned
symmetry. We support this novel approach by presenting a first
successful Hubbard-I type calculation for the description of the
metal-insulator Mott transition in a strongly correlated electron
system with conserved double occupancies, which is a constrained
Hubbard Hamiltonian equivalent to the Hubbard bond charge Hamiltonian
with $X=t$.  In particular, we obtain the phase diagram of this system
for arbitrary fillings, including details of the Mott transition at
half-filling. We also compare the Hubbard-I band--splitting Mott
transition description with results obtained using the standard
Gutzwiller Approximation (GA), and show that the two approximate
approaches lead to qualitatively different results. In contrast to the
GA applied to the system studied here, the Hubbard-I approach
compares favourably with known exact results for the $d=1$ dimensional
chain.

\end{abstract}

 
 \maketitle

\section{Introduction}

  The metal to Mott insulator (MI) transition, envisaged by Mott
  \cite{Mott}, is one of the striking effects induced by strong
  electronic correlations in many electron systems.  The Coulomb
  interaction between electrons $U$ leads to highly correlated ground
  states in these systems, rendering their description quite
  challenging.  As a result there have been various theoretical
  attempts at providing a satisfying description of the Mott
  transition.

The first of these was due to Hubbard \cite{HubbardI}, who provided a
seminal nonperturbative approach -- the so-called Hubbard-I (H-I)
approximation -- to a simplified interacting electron problem
described by the Hubbard model. H-I describes both (i) the atomic
limit ({\it i.e.} limit of vanishing bandwidth $W \rightarrow 0$)
exactly, in particular yielding two atomic levels corresponding to
single and double local occupancy, and (ii) the non-interacting case
($U=0$) exactly, and so held some promise of describing the
intermediate physics in a consistent interpolating fashion.  For
finite bandwidth, the atomic levels broaden into two ``dynamic''
(sub)bands which are occupation-number and interaction
dependent. These are always split by a gap for all $U>0$ indicating an
insulator phase.  Unfortunately H-I approach in relation to the Hubbard model 
is flawed at a basic level. (i) It is not a particle-hole symmetry conserving approximation
  \cite{HubbardI} (it is not guaranteed that the Mott insulator phase exists only exactly at half-filling).  (ii) Hubbard-I predicts a Mott transition at
  $U=0$, which certainly is not true in general - the Hubbard model on
  the honeycomb lattice {\it e.g.} has a finite critical point $U_{c}$
  \cite{Paiva2005}.  More importantly, the Hubbard-I does not yield a
  viable description of the weak coupling limit $U/W \ll 1$, in
  particular it does not reproduce the renormalized Fermi liquid
  expected from the Hartree-Fock approximation.
The situation
was somewhat improved in the Hubbard-III approximation
\cite{HubbardIII}, where scattering and resonance broadening
corrections shift the transition point to $U_{c}\sim W$. However, this
approach does not predict the expected Fermi liquid properties on the
metallic side, \cite{Edwards}. The Hubbard approaches were very important for the conceptual introduction of the 
Hubbard subbands, however their problems have led some autors to the rejection of Hubbard approximations for the study
  of strongly correlated electron systems.  In spite of these issues,
  however, as shown in a recent study by Dorneich {\it et al.}
  \cite{Dorneich2000}, a suitably generalized Hubbard-I approach
  actually gives reasonably good description even at the quantitative
  level in the limit of strong interaction (large $U$) and weak spin
  correlations, of the Hubbard model.

A complementary approach, starting from the weak coupling limit, is
based on the Gutzwiller variational wave function \cite{Gutzwiller}
(GWF), which describes an increasingly correlated Fermi liquid with
increasing interaction $U$. Brinkman and Rice \cite{Rice} showed,
using the so-called Gutzwiller approximation \cite{Gutzwiller} (GA),
that increase in $U$ is associated with the diminishing of the
quasiparticle residue of the Fermi liquid, or equivalently with the
increase in quasiparticle effective mass. In this framework, a
metal-insulator transition occurs at finite value $U_{c}$ when the
effective mass diverges and is thus driven by quasiparticle
localisation.  This method gives a good low energy description of the
metal, but does not describe the precursors of the Hubbard bands which
should plausibly appear on the metallic side. Importantly, for finite
dimensional systems, the GBR transition is an artefact of the GA,
since analytical \cite{Millis,Dzierzawa} and numerical studies
\cite{Yokoyama} of the GWF show that it always describes a metallic
state.

Although, both these early methods have their flaws, they hint at two
possible mechanisms behind the Mott transition. These two pictures
have been brought to the forefront, more recently, with the
development of more involved contemporary methods, in particular
dynamical mean field theory (DMFT) \cite{Kotliar} which provides a
bridge between the formation of Hubbard bands on the one hand, and
strongly correlated fermi liquid behaviour on the other. In
particular, the DMFT inspired modern prevailing view is that
quasiparticle localization drives the ground state Mott transition in
the Hubbard model \cite{Kotliar}.

In this paper, we reconsider the Hubbard-I type
approach as a systematic means of obtaining a band-splitting theory of
the Mott transition. Instead of trying to improve H-I approximation as a method for general strongly correlated Hamiltonians we rather seek to specify which Hamiltonians may be successfully studied by a standard H-I approximation.
We shall argue that there are correlated Hamiltonians, {\it viz.}
  those describing electron systems with so-called extreme
  correlations \cite{Shastryex}, for which the Hubbard-I approach
provides a basic "mean-field"--like description.

The paper is organized as follows: In Section 2 we discuss the relevance of  double occupation conservation symmetry for the application of the H-I approximation. In Section 3 we consider the H-I approach to one of the simplest Hamiltonians preserving this symmetry as a model of a band-splitting theory of the Mott transition. In Section 4 we compare  results with known features of the Hubbard bond-charge model - an equivalent model on bipartite lattices. Section 5 consists of a short discussion of corrections to the H-I approach for the studied Hamiltonian. Finally section 6 is devoted to a comparison of two complementary theories of the Mott 
transition:  band-splitting driven and quasiparticle localisation driven.

\section{Inspection of the Hubbard-I approach}

Contrary to early expectations, we now see the Hubbard approach, based on a
Green's function decoupling scheme, as a large--U
approximation, despite its non perturbative nature. One may therefore
suspect that some problems encountered in the Hubbard approach may
originate in limitations of strong-coupling perturbation theory.

Consider therefore the Hubbard model: $H= \hat{T} + U\hat{D}$, where
$\hat{D}= \sum_{i} n_{i \uparrow} n_{i \downarrow}$ is the number of
doubly occupied sites and $ \hat{T}=\sum_{i,j, \sigma }^{}
t_{i,j}\fer{i}{\sigma
  } ^{\dagger}\fer{j}{\sigma} $ is the kinetic energy. The large-U
perturbational expansion starts from splitting the kinetic energy into
three parts $ \hat{T}= \hat{T}_{0}+ \hat{T}_{+1} +
\hat{T}_{-1}$. $\hat{T}_{0} = \hat{T}_{UHB} + \hat{T}_{LHB}$ is the
double occupancy conserving hopping (commuting perturbation) in the
Upper and Lower Hubbard Bands:
\begin{eqnarray}
  \hat{T}_{UHB}= \sum_{i,j, \sigma }^{} t_{ij}n_{i \sigmab } \fer{i}{\sigma
  } ^{\dagger}\fer{j}{\sigma} n_{j \sigmab  } ,
\nonumber \\  
   \hat{T}_{LHB}= \sum_{i,j, \sigma }^{} t_{ij}(1-n_{i \sigmab }) \fer{i}{\sigma
  } ^{\dagger}\fer{j}{\sigma} (1-n_{j \sigmab  }),
\nonumber
\nonumber 
\end{eqnarray}
corresponding to hopping of projected electrons on doubly and singly
occupied sites respectively. The remaining hopping terms
  $\hat{T}_{+1} $ and $ \hat{T}_{-1}=\hat{T}_{+1}^{\dagger}$
  correspond to interband hopping (non--commuting perturbation).  The
perturbational expansion for the Hubbard model is well known and can
be performed e.g. using the method of canonical transformations
\cite{HarrisLange,Chao} which to second order yields the effective
Hamiltonian:
\begin{equation}
H^{(2)}_{\rm eff} = \hat{T}_{0} + U\hat{D} +\frac{1}{U}  [ \hat{T}_{+1},
\hat{T}_{-1} ].
\label{constrained}
\end{equation}

Recall that the perturbation expansion to any order  eliminates mixing between the degenerate
subspaces of the unperturbed Hamiltonian $\hat{U}$ and  leads to an
effective Hamiltonian with emergent symmetry of conservation of the number of
doubly occupied sites. Due to this
property, the average of operators jointly changing the total number
of doubly occupied sites, such as {\it e.g.} $\langle \fer{j}{\sigma
}n_{j \sigmab } c_{l \sigma }^{\dagger}h_{l \sigmab
}\rangle$,($h_{j\sigmab}=1-n_{j\sigmab}$ is the on-site hole
occupation number), are identically zero leading to the vanishing of
the associated Green's functions. Thus in this framework, the single
particle Green's function $G_{lj}^\sigma=\langle\langle \fer{l}{\sigma
}| c_{j \sigma }^{\dagger} \rangle\rangle$ decomposes into a sum of
two Green's functions $\Gamma_{lj}^{\sigma } = \langle\langle
\fer{l}{\sigma }n_{l \sigmab } | c_{j \sigma }^{\dagger}n_{j \sigmab }
\rangle\rangle$ and $\widetilde{\Gamma}_{lj}^{\sigma } =
\langle\langle \fer{l}{\sigma }h_{l \sigmab } | c_{j \sigma
}^{\dagger}h_{j \sigmab } \rangle\rangle$ related to the propagation
of fermionic quasiparticles, in the upper and lower Hubbard bands respectively.

Interestingly the Hubbard-I approach is most 
naturally described by the separation of an electron into two
non-canonical fermions $\fer{j}{\sigma }=\fer{j}{\sigma }n_{j
  \sigmab}+\fer{j}{\sigma }h_{j \sigmab}$ \cite{HubbardIV}. Consider therefore the H-I approach to the simplest non-trivial level of
strong-coupling perturbational expansion, described by the first two
terms in the expansion Eq.(\ref{constrained})
\begin{equation}
H_{c}=\hat{T}_{0} +U \hat{D}. 
\label{hamconstrained}
\end{equation}
In the frequency domain, the equations of motion for the basic Green's
function are $\omega \Gamma_{lj}^{\sigma }(\omega ) = \frac{1}{2 \pi
}\langle n_{l \sigmab } \rangle \delta_{lj} + \langle\langle[
  \fer{l}{\sigma } n_{l \sigmab},H_{{\rm c}}]| \fer{j}{\sigma
}^{\dagger} n_{j \sigmab } \rangle\rangle_{\omega } $ and $\omega
\widetilde{\Gamma}_{lj}^{\sigma }(\omega ) = \frac{1}{2 \pi } \langle
h_{l \sigmab } \rangle \delta_{lj} + \langle\langle[ \fer{l}{\sigma }
  h_{l \sigmab},H_{{\rm c}}]| \fer{j}{\sigma }^{\dagger} h_{j \sigmab
} \rangle\rangle_{\omega } $. We perform the following mean-field or H-I type of
decoupling on the higher order Green's function in the equations for
$\Gamma_{lj}^{\sigma }(\omega )$ to terminate the sequence of equations at the level of the upper and lower band Green's functions:
\begin{eqnarray}
 \langle\langle \fer{l + \delta }{\sigma } n_{l+ \delta \sigmab} n_{l \sigmab}  | \fer{j}{\sigma
  }^{\dagger} n_{j \sigmab }\rangle\rangle_{\omega } \approx \langle n_{l
    \sigmab}\rangle \Gamma_{l+ \delta ,j}^{\sigma } (\omega )
\nonumber 
  \\
 \langle\langle\fer{l}{\sigma }\fer{l}{\sigmab } ^{\dagger}\fer{l +
  \delta }{\sigmab } n_{l+ \delta \sigma} | \fer{j}{\sigma
}^{\dagger} n_{j \sigmab }\rangle\rangle_{\omega } \approx \langle \fer{l}{\sigma
  }\fer{l}{\sigmab } ^{\dagger} \rangle \langle\langle \fer{l +
    \delta }{\sigmab } n_{l+ \delta \sigma} | \fer{j}{\sigma
  }^{\dagger} n_{j \sigmab }\rangle\rangle_{\omega }
\nonumber 
\\
    \langle\langle h_{l + \delta \sigma } \fer{l+ \delta
}{\sigmab } ^{\dagger} \fer{l}{\sigmab } \fer{l}{\sigma } | \fer{j}{\sigma
  }^{\dagger} n_{j \sigmab }\rangle\rangle_{\omega } \approx  \langle \fer{l}{\sigmab
  }\fer{l}{\sigma } \rangle \langle\langle \fer{l +
    \delta }{\sigmab }^{\dagger} h_{l+ \delta \sigma} | \fer{j}{\sigma
  }^{\dagger}n_{j \sigmab } \rangle\rangle_{\omega } 
\nonumber 
\end{eqnarray}
 with analogous complementary treatment of the set
 $\widetilde{\Gamma}_{lj}^{\sigma }(\omega )$. The last two
 decouplings lead to magnetic and superconducting order parameters and
 which are  set here to zero, as we are interested in the description of a
 paramagnetic Mott transition. Solving the equations of motion, one
 obtains the following momentum (or Fourier transformed) Green's
 functions for the system:
\begin{eqnarray}
\Gamma_{\bm{k}}^{\sigma } (\omega )= \frac{1}{2 \pi }\frac{n_{\sigmab }}{\omega
  - (n_{\sigmab } \epsilon_{\bm{k}} +U - \mu )}
\label{R6ab}
\\
\widetilde{\Gamma}_{\bm{k}}^{\sigma } (\omega )= \frac{1}{2 \pi }\frac{1- n_{\sigmab }}{\omega
  - ((1-n_{\sigmab }) \epsilon_{\bm{k}} - \mu )}
\label{R5ab}
\end{eqnarray}

The electron Green's function $ G_{\bm{k}}^{\sigma } (\omega )=\langle\langle \fer{k}{\sigma} | \fer{k}{\sigma}^{\dagger} \rangle\rangle_\omega= \Gamma_{\bm{k}}^{\sigma } (\omega )+\widetilde{\Gamma}_{\bm{k}}^{\sigma } (\omega )$ thus has the characteristic Hubbard-I two
pole form describing two subbands, however the poles here have an
elementary form in comparison with the solution for the full Hubbard
model \cite{HubbardI}. The main result stemming from these Green's
functions is that a Mott transition, identified with band separation, is elevated to finite interaction
$U$. Indeed, for a paramagnetic configuration at half filling $\langle
n_\uparrow\rangle=\langle n_\uparrow\rangle=1/2$ the lowest energy in
the UHB is readily read out from Eq.(\ref{R6ab}) to be $U-W/4$, while the highest energy in the LHB is from Eq.(\ref{R5ab}) $W/4$ (where
$W$ is the free electron bandwidth), hence a gap opens at a critical
value $U_c=W/2$. 
Importantly, the
presented solutions Eqs.(\ref{R6ab},\ref{R5ab}) are explicitly
electron hole symmetric.
Accordingly, the gap opening at $U=W/2$ is
associated with a Mott transition only exactly at half-filling.

These results provide explicit evidence that the source of problems in
the Hubbard-I approach stem from the effects of the double occupancy
non--conserving terms $T_{\pm 1}$.  Since the Hubbard-I approach leads to the separation of an electron into two
non-canonical fermions $\fer{j}{\sigma }=\fer{j}{\sigma }n_{j
  \sigmab}+\fer{j}{\sigma }h_{j \sigmab}$ which are subsequently
  treated separately, 
  we view the source problem as that of
incompatibility of the one band electron normal Landau-Fermi liquid
(consistent with the full hopping operator $\hat{T}$) with the two
(bands of) non-canonical fermion liquids (consistent with double
occupancy number conservation). In fact, the two non-canonical
liquids cannot be adiabatically connected to the normal Fermi liquid,
so one should rather consider the Hubbard-I as a mean-field type
approach appropriate for systems conserving double occupancies, such
as those obtained in strong coupling perturbational theory at any
order, not arbitrary systems. We defer discussion of our
Hubbard-I approach to other and higher order Hamiltonians to
later papers, and in the remainder of this paper we shall analyze
results for the model in Eq.(\ref{hamconstrained}).

\section{The Mott transition}

The constrained Hamiltonian $H_c$ Eq.(\ref{hamconstrained}) is
interesting in itself because it can be viewed as a particular model
of extremely correlated electrons. Recall that the term extremely
correlated electron systems has recently been introduced by Shastry
\cite{Shastryex}, emphasizing the appearance of noncanonical fermions
resulting from the prohibition of double occupancies in the $U=\infty$
limit, which is a special case of symmetry of conserved double
occupancies.  The model $H_c$ considered here, that allows for double
occupancies which are conserved and thus describes hopping of
noncanonical (constrained) fermions $\fer{j}{\sigma} n_{j\sigmab}$ and
$\fer{j}{\sigma} h_{j\sigmab}$ can therefore be considered a
generalization of the problem of extremely correlated fermions for
general interaction strength $U$. Note that the model $H_c$ contains
the $U \rightarrow \infty$ limit of the Hubbard model due to its perturbative origin. Even for small
$U$ and low fillings, we discuss below that the ground states of $H_c$
correspond to the ground states of the $U=\infty$ Hubbard model.

We now consider the details of the Hubbard-I description of the zero
temperature Mott transition in the model $H_c$ of extremely correlated
electrons,  as it is a first successful theory of a band-splitting driven Mott transition. The analysis is carried out only for the paramagnetic
phases, with $n_\uparrow=n_\downarrow$.

\begin{figure*}[htb]
  	\centering
    \includegraphics[scale=0.5]
    {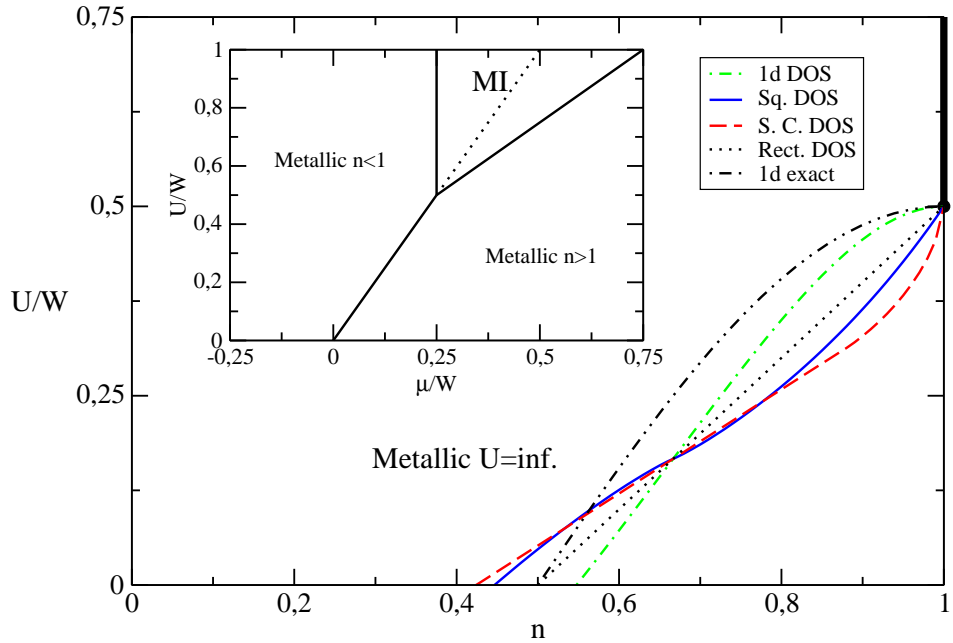}
    \caption{Ground state phase diagram of the model $H_c$ in the
      $U-n$ plane. In all calculations we assume only nearest
      neighbour hopping $t_{ij}=-t$. A metallic ECFL exists for all $n
      \leq 1$. At low filling, the metal resembles the $U=\infty$
      metallic phase of the Hubbard model, with no double
      occupancies. Closer to $n=1$, the metal has double occupancies
      and is described by two Fermi liquids per spin species. In the
      H-I calculation, the boundary between these two behaviours is a
      Lifshitz-type transition, shown for hyper-cubic lattices in 1,2,3
      dimensions and for the rectangular DOS. Shown is also the exact
      result for 1d, from \cite{Ovchinnikov} with rescaled
      $U\rightarrow U/2$, for comparison. All transition lines cross
      the $n=1$ line at finite $U=W/2$ marking the point of Mott
      transition. Solid line depicts the Mott phase. The $n>1$ can be
      obtained by invoking electron-hole symmetry. Inset: Ground state
      phase diagram in the $U-\mu$ plane. The electron hole symmetry
      line (dotted) separates $n<1$ and $n>1$ cases. The $n=1$ case
      coincides with this line for $U<W/2$ whereas for $U>W/2$ opens
      into a Mott insulator "lobe" of finite width.}
    \label{fig1}
  \end{figure*}

From the band structure of the one particle Green's function, it is
evident that the insulator phase exists only if $U>U_c$ and the lower
band is completely filled while the upper band is empty. The Green's
functions Eqs.(\ref{R6ab},\ref{R5ab}) lead to the following equation
for the number of electrons of a given spin species:
\begin{eqnarray}
n_{\sigma}=(1-n_{\sigmab}) \int_{-W/2}^{W/2}\! 
f((1-n_{\sigmab})\epsilon -\mu)\rho (\epsilon) \,
\mathrm{d}\epsilon
 \nonumber\\
+ n_{\sigmab} \int_{-W/2}^{W/2}\! 
f(n_{\sigmab}\epsilon +U -\mu)\rho (\epsilon) \,
\mathrm{d}\epsilon
\label{nsigma}
\end{eqnarray}
where $f(\ldots)$ is the Fermi-Dirac distribution and $\rho
(\epsilon)$ is the density of states. The boundaries of the insulator
phase are obtained, independently of the lattice, when either $\mu=
(1-n_{\sigmab}) W/2$ is at the end of the lower band or $\mu =
-n_{\sigmab} W/2 + U$ is at the beginning of the upper band. For both
these cases, at zero temperature, Eq.(\ref{nsigma}) reduces to
$n_{\sigma}=1-n_{\sigmab}$, showing that the transition only occurs at
half-filling, as claimed. The jump in $\mu$ at half-filling $n_\uparrow =
n_\downarrow=1/2$ reduces to zero at $U=W/2$ indicating the transition
point.  The Mott ``lobe'' is shown in Fig.\ref{fig1}. Note that the
kinetic energy and interaction energy are both zero in the Mott phase,
which is an exact feature of the model. Indeed, note that there
  is extensive degeneracy of localized Mott states consisting of one
  particle per site which are all exact ground states of the
  considered model (Eq.(\ref{hamconstrained})), with zero kinetic
  energy due to double occupancy conservation.

Outside the Mott phase, the system is in a metallic phase described,
in the used approximation, by four (two per spin species)
dispersion-less extremely correlated Fermi liquids (ECFLs) associated
with the lower and upper Hubbard bands
Eq.(\ref{R6ab},\ref{R5ab}). Depending on the filling factor and value
of $U$, for each spin species one can obtain a situation where only
one kind of Fermi liquid of $\fer{j}{\sigma } h_{j \sigmab }$ fermions
appears, to which there pertains a single well defined Fermi surface
(see Fig. \ref{fig1}). For $n<1$, this case is physically equivalent
to the infinite $U$ limit of the Hubbard model.  The single Fermi
liquid comes about because for small $n_\uparrow,n_\downarrow$ the
lower Hubbard band is wide while the upper Hubbard band is very
narrow. Even for $U<W/2$, although the two bands overlap, the upper
band centered around $U$ is too narrow to be occupied in the ground
state. On the other hand, closer to half filling $n\leq 1$ for $U<W/2$
when the upper band can also be occupied, two coexistent Fermi liquids
of $\fer{j}{\sigma } h_{j \sigmab }$ and $\fer{j}{\sigma } n_{j
  \sigmab }$ fermions emerge, per spin species, associated with two
Fermi surfaces. The boundary between these two cases is therefore a
Fermi surface topology changing transition, which can be called a
Lifshitz type transition.

The above band picture comes with a caveat, {\it viz.} the bands
Eq.(\ref{R6ab},\ref{R5ab}) which seem to be apparently independent in
the Grand Canonical Ensemble (GCE) cannot be treated as such when
attempting to construct arbitrary excitations. For constructing
elementary excitations, at the very least an additional rule must be
implemented to obtain physically relevant states, {\it i.e.} the upper
bands pertaining to both spin indices must be equally populated (these
are equal to the double occupancy) while the number of holes in the
lower bands must also be equal. This may be seen as a type of
``statistical interaction'' \cite{Haldane,Gebhard} between the bands
in this system.  In the GCE, this problem is masked and taken care of,
by the common value of the chemical potential for both species which
guarantees equal population of the upper band and equal number of
holes in the lower band, in the thermodynamic limit.

\begin{figure*}[htb]
  	\centering
    \includegraphics[scale=0.5]
    {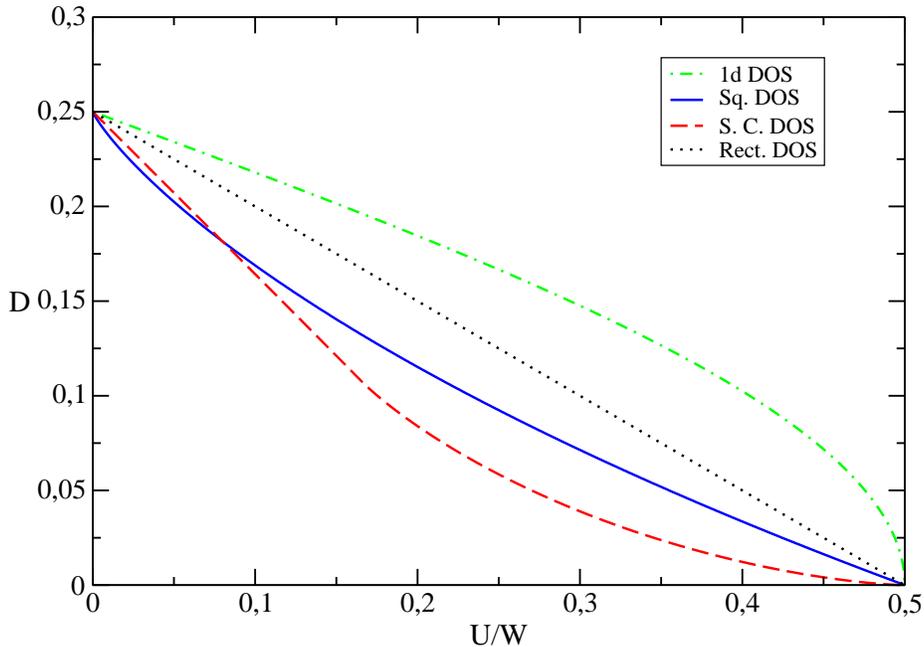}
    \caption{The double occupancy per site $D$ at half-filling $n=1$
      as a function of $U$. Shown are results for hyper-cubic lattices
      in 1, 2, 3 dimensions and for rectangular DOS. For all DOSes
      $D=1/4$, at $U=0$, as in the uncorrelated metal, although it is
      a ECFL here.  The Mott transition takes place at $U=W/2$. Note
      different critical behavior in 1,2 and 3 $d$. The curve
      calculated for 1$d$ coincides with known exact result
      \cite{Ovchinnikov} for rescaled $U\rightarrow U/2$.}
    \label{fig2}
  \end{figure*}

We now focus on the interaction driven Mott transition from the
metallic phase.  Unlike in the Gutwziller method, the hopping in the
approach discussed here cannot be used as an indicator of the
transition, as it remains constant.  The density of doubly occupied
sites $D=\langle \hat{D} \rangle/N$ (occupation of the upper Hubbard
band) is a good parameter at half filling capturing the transition for
this model. Using the Green's function of Eq.(\ref{R6ab}) (or
  the second term in Eq.(\ref{nsigma})) and the half-filling condition
  $\mu=U/2$, one obtains the particularly simple form at zero
  temperature
\begin{equation}
D=\langle \hat{D} \rangle/N=\frac{1}{2}\int_{-W/2}^{-U}\! 
\rho (\epsilon) \,
\mathrm{d}\epsilon
\label{doubles}
\end{equation}
which depends only on $U$ and the density of states. We show results for
a few different density of states (DOSes) in Fig.\ref{fig2}.

Note that an important feature of our Hubbard-I calculation is that
the critical point does not depend on the lattice type.
 Furthermore, lattice dependent properties emerge for the number of
 doubly occupied sites. A linear dependence is associated here only
 with a rectangular density of states. In fact, the critical behaviour
 of $D$ has universal properties depending only on spatial dimensions
 of the lattice. It is governed by the behaviour of the DOS at the
 bottom of the band. Indeed for parabolic energy dispersions near the
 bottom of the band, as ${\epsilon}\rightarrow -W/2 \Rightarrow
 \Delta\epsilon_{\bm{k}} \propto {\bm{k}}^2$, we have for 1 dimension
 $\rho(-W/2+\Delta\epsilon)\propto 1/\sqrt{\Delta\epsilon} $, 2
 dimensions that $\rho(-W/2+\Delta\epsilon)\propto \mathrm{const.} $,
 and 3 dimensions $\rho(-W/2+\Delta\epsilon)\propto
 \sqrt{\Delta\epsilon} $. Then we obtain that in 1d: $ D \propto
      [(U_c-U)/{U_c}]^{\frac{1}{2}}$ , 2d: $ D \propto (U_c-U)/{U_c}$
      while in 3d: $ D \propto [(U_c-U)/{U_c}]^{\frac{3}{2}}$.

\section{ Relation to the Bond-charge Hubbard model}

While the results presented in the previous paragraph provide a clear
and simple quantitative picture of a paramagnetic transition as well
as the metallic phase, one may wonder if the constrained model $H_c$
is realistic, as it was derived as lowest order of strong-coupling
expansion. Additionally, it is not a priori clear how good an
approximation the Hubbard-I approach yields for the constrained model
$H_c$. Therefore it is important to note that this model is related to
the class of generalized Hubbard models, differing from the Hubbard
Hamiltonian by an additional bond charge interaction term (see {\it
  e.g.}:
\begin{equation}
H_{\rm bond-charge} = \hat{T} + U \hat{D} + \sum_{i,j,\sigma} X_{ij}
(n_{i \sigmab }+n_{j \sigmab } )\fer{i}{\sigma }
^{\dagger}\fer{j}{\sigma}
  \label{bondcharge}
\end{equation}
which was already discussed by Hubbard \cite{HubbardI} and
reconsidered in relation to superconductivity by
\cite{Micnas,Hirsch}). Interestingly this is  one of the few models
in which there is a Mott transition in the ground state, for certain
values of parameters \cite{Strack}.

Indeed, at a symmetry point, when the bond charge interaction is equal
to the hopping $X_{ij}= -t_{ij}$, this generalized Hubbard model
conserves the number of double occupancies. In fact, the resultant
symmetric model can be mapped onto the model $H_c$ on bipartite
lattices, via the canonical transformation $U=\exp [-i \pi
  \sum_{j}{\bm R}_j n_{j \uparrow} n_{j \downarrow} ]$, where $R_j =
\pm 1$ on different sublattices, as shown in
\cite{Gebhard}. Analytical ground states of the symmetric bond-charge
model were obtained in \cite{Strack} in the regime $U>2W$ at half
filling, for arbitrary dimensions $d$. These results were improved
upon in \cite{Ovchinnikov} revealing that a metal-insulator transition
occurs at $U=W$, probably with ferromagnetic polarization on the
metallic side (for $d>1$), which is supported by numerical evidence
\cite{Gagliano}.

The special case of $d=1$, was exactly solved in \cite{Ovchinnikov}
leading to the following main results: the metal to insulator
transition occurs at $U_c=W$, and also the number of double
occupancies is $D= 1/2 \pi {\rm ArcCos}(U/U_c)$.  In $d=1$, the H-I
approach used here grossly underestimates the transition point, but
remarkably the average number of double occupancies calculated for the
1-d DOS $\rho(\epsilon)= 1/(\pi \sqrt{1-\epsilon^2})$, using
Eq.(\ref{doubles}) leads to exactly the same function and prefactor as
the exact solution (with $U_c= W/2$ now being the H-I critical value).
Interestingly, the exact results can also be interpreted in terms of
lower and upper Hubbard bands, as shown in \cite{Gebhard}. These bands
however do not carry spin indices and the hopping is not renormalized,
while also a different mechanism of statistical interaction than the
one obtained here is present.

 Furthermore, exact results away from half-filling $n=1$
reveal, that there is a critical boundary separating states with
doubly occupied sites from states with no such sites, which in the H-I
framework considered here corresponds to the Lishitz lines $U_L$
depicted in Fig.\ref{fig1}. The exact boundary is given by the
relation $U_L = W \cos (\pi (1-n))$ \cite{Ovchinnikov}. Notice that,
apart from the value of the Mott transition point, the Lifshitz line
obtained using the H-I approach here is in good quantitative agreement
with these results (see Fig.\ref{fig1}). In general dimensions, the
expected phase diagram is expected to be qualitatively similar
\cite{Schadschneider} as in $1$ dimension. We thus see, that the phase
diagram derived by the Hubbard-I approach is in good qualitative
agreement with exact results.

\section{The Roth--corrected approach}

The critical point $U_c$ in the Hubbard-I approach is grossly
underestimated and additionally, for general dimensions, does not
depend on magnetic polarization, since the gap is always given by
$U-W/2$ (see Eqs. (\ref{R6ab}, \ref{R5ab})). In this subsection we
indicate that these drawbacks may be
 treated in a unified manner by the procedure invoked by Roth
 \cite{Roth}, which removes ambiguities in the decoupling of Green's
 functions equations of motion method. Applying the Roth procedure at
 the Hubbard-I level of equations of motion for the considered
 Hamiltonian $H_c$, we obtain the following Green's functions:
\[
\Gamma_{\bm{k}}^{\sigma } (\omega )= \frac{1}{2 \pi }\frac{n_{\sigmab
}}{\omega - E_{\sigma}} , \;
\widetilde{\Gamma}_{\bm{k}}^{\sigma } (\omega )= \frac{1}{2 \pi
}\frac{h_{\sigmab }}{\omega - \widetilde{E}_{\sigma}}
\]
where, the quasi-particle energy factors are
\begin{eqnarray}
E_{\sigma} = (U-\mu) + \epsilon_{\bm{k}} \langle n_{\bm{0} \sigmab}n_{
  \bm{\delta}\sigmab}\rangle/n_{\sigmab} \nonumber
\\ -\epsilon_{\bm{k}} (\langle T_{D} \rangle + \langle T_{X}
\rangle)/n_{\sigmab} + (\langle T_{ UHB}\rangle - \langle T_{
  LHB}\rangle)/n_{\sigmab},
\label{green1}\\
\widetilde{E}_{\sigma} = -\mu + \epsilon_{\bm{k}} \langle h_{\bm{0}
  \sigmab}h_{ \bm{\delta}\sigmab}\rangle/h_{\sigmab} \nonumber
\\ -\epsilon_{\bm{k}} (\langle T_{D} \rangle + \langle T_{X}
\rangle)/h_{\sigmab} + (\langle T_{ UHB}\rangle - \langle T_{
  LHB}\rangle)/h_{\sigmab}.
\label{green2}
\end{eqnarray}
The term $\langle T_{X} \rangle \equiv \langle \fer{\bm{0}}{
  \sigmab}^{\dagger} \fer{ \bm{\delta}}{\sigmab}\fer{
  \bm{\delta}}{\sigma}^{\dagger} \fer{\bm{0}}{ \sigma} \rangle$
describes the process of interchange of spins between neighbouring
sites, while $\langle T_{D} \rangle \equiv \langle\fer{\bm{0}}{
  \sigmab}^{\dagger} \fer{ \bm{\delta}}{\sigmab}\fer{
  \bm{0}}{\sigma}^{\dagger} \fer{\bm{\delta}}{ \sigma} \rangle$
describes the process of doublon transfer. The averages in
Eqs. (\ref{green1}, \ref{green2}) are yet to be determined quantities,
which is a standard feature of Roth's method.  Notice, that
disregarding the second lines in Eqs. (\ref{green1}, \ref{green2}),
and considering site occupations as uncorrelated, one recovers the
Hubbard I results.

A full analysis of the Roth solutions shall be considered elsewhere
\cite{Grzybowski}, while here we only consider implications for the
Mott phase,  for which $\langle T_{X} \rangle, \langle T_{D} \rangle
,\langle T_{ UHB}\rangle, \langle T_{ LHB}\rangle $ vanish identically
for the considered model $H_c$.  The closure of the spectral gap to
excitations indicates the Mott phase instability.  Consider the
half-filled case with $n_\uparrow=n_\downarrow=1/2$. If there are no
inter-site correlations in the Mott states, $U_c=W/2$ as indicated
above. On the other hand for saturated ferromagnetic correlation, {\it
  i.e.} macroscopic separation of the system into two oppositely
polarized ferromagnetic domains $\langle n_{\bm{0} \sigmab}n_{
  \bm{\delta}\sigmab}\rangle =1/2 = n_{\sigmab } =\langle h_{\bm{0}
  \sigmab}h_{ \bm{\delta}\sigmab}\rangle = h_{\sigmab } $ ,
Eqs. (\ref{green1}, \ref{green2}) show that there is no
renormalization of the band width in the Hubbard bands, and thus
$U_c=W$. As all Mott states have the same energy, this analysis
indicates that the Mott phase is unstable already at $U_c=W$ (which is
in good agreement with known results summarized earlier), and changes
probably into a ferromagnetic metallic state.

\section{Comparison with the Gutzwiller approach}

It is worthwhile to compare, for completeness, our H-I results, for
half filling, with those obtained using the Gutzwiller approximation
to the model $H_c$.  We perform a standard calculation (see {\it e.g.}
\cite{Vollhardt}), {\it i.e} assuming the reference state, that is
Gutzwiller projected, to be the product of Fermi seas of the two spin
species.  Within the GA, the average energy of the Hamiltonian $H_c$
can be written as:
\[
\langle H_c \rangle = \sum_{\sigma} q_{\sigma} \epsilon_{0 \sigma} + UD, 
\]
where $\epsilon_{0 \sigma}$ is the energy of the Fermi sea of a given
spin species. The band narrowing factor $q_{\sigma}$ is easily found
to be given by:
\[
q_{\sigma} = \frac{(n_{\sigma}- D)(1+D-n_{\sigma} - n_{\sigmab})
  +D(n_{\sigmab}-D)}{n_{\sigma}(1-n_{\sigma})}.
\]
Minimizing the average energy with respect to $D$, at half filling, we
obtain a Gutzwiller-Brinkman-Rice transition(GBR) at a finite $U_c =
-4 |\epsilon_0|$, where $\epsilon_0$ is the summed ground state energy
of the noninteracting Fermi liquids. The GA density of doubles on the
metallic side is linearly dependent on the interaction $U$, reducing
to zero at the transition point:
\[
D= \frac{1}{4} \Big( 1- U/U_c\Big), \;\;\; U\leq U_c . 
\]

\begin{table*}
 \caption{\label{coll}Comparison of Gutziller Approximation (GA) and
   Hubbard-I (H-I) results with known exact results \cite{Ovchinnikov,Schadschneider}
   for the bond-charge model at the symmetry point $X=t$.}
 \begin{tabular}{@{}llll}
 \hline
  & GA & H-I& Exact (X=-t)\\
 \hline
 $U_{c}$ &  lattice dependent & lattice independent & lattice
 independent \\
 &&$U_{c}=W/2$& $U_{c}=W$\\
 $D(U)$&lattice independent& lattice dependent & \\
 &$D=\frac{1}{4}(1- U/U_{c})$ & $D= \frac{1}{2}\int_{-U_{c}}^{U}
 \rho (\epsilon )d \epsilon $ & \\
 & & (1d) $D= 1/2 \pi {\rm ArcCos}(U/U_c)$   & (1d)  $ D= 1/2 \pi {\rm ArcCos}(U/U_c)$  \\
 $D(U) \sim (\frac{U-U_{}}{U_{c}})^{\beta }$&lattice independent& lattice dependent&\\
 & $\beta =1$ & (1d) $\beta =1/2$ & (1d) $\beta =1/2$\\
 &  & (2d) $\beta =1$ & \\
 &  & (3d) $\beta =3/2$ & \\
 \hline
 \end{tabular}
 \end{table*}

The GBR transition is of a quantitatively different character than the
obtained H-I transition. Indeed, the distinct points are that the
transition point is lattice dependent, while the metallic side has
lattice independent double occupancies (which seem to be doubtful in
the light of some exact results summarized above). As an aside, note
however that, the choice of the reference state as noninteracting
uncorrelated Fermi seas is of doubtful applicability, in the
Gutzwiller analysis, of the model $H_c$ given that it certainly does
not even correspond to the ground state of a correlated hopping
Hamiltonian anywhere, including in the limit $U=0$.  This is analogous
to the Gutzwiller study of the bond-charge Hamiltonian
Eq.(\ref{bondcharge}) performed in \cite{Kollar}, where the results
for large bond charge interaction $X$ cannot be considered
reliable. Indeed, in particular at the symmetry point (which is of
direct relevance to the model $H_c$), $X_{ij}= -t_{ij}$, Kollar and
Vollhardt in \cite{Kollar} obtain a critical point $U_c=0$ using the
GA. Thus, rather interestingly, we observe that the Gutzwiller
approximation yields a qualitatively better picture of the phase
transition itself in the model $H_c$ than in the bond-charge
Hamiltonian. 

 Finally, the GA shows a ferromagnetic instability in the bond-charge
 model, for certain lattices \cite{Kollar} before the Mott insulator
 transition. However, this is not the case for the GA in the $H_c$
 Hamiltonian. Indeed, as usual \cite{Kollar,Rice} one can calculate
 the bulk magnetic susceptibility $\chi$, which is here given by:
\[
\chi^{-1} = \frac{1-(U/U_c)^2}{2\rho(\varepsilon_F)}(1- \rho(\varepsilon_F)U_c)
\]
where $\rho(\varepsilon_F)$ is the density of states at the Fermi
surface.  Only one factor depends on $U$ and can diverge here,
accompanying the metal insulator transition $U \rightarrow U_c$. Thus,
the GA, like the H-I calculation, does not describe a ferromagnetic
metal before the Mott transition.  However, the phase boundaries
  and metallic properties are better described by the H-I approach, as
  seen on comparison with exact results recalled in Table \ref{coll}.

\section{Conclusions and Outlook}

In this work, we have analyzed the Hubbard I approximation, explicitly
showing that its known drawbacks originate from the interband hopping
$T_{\pm 1}$ terms in the Hubbard model. We proposed to use the H-I
approach in conjunction with perturbational expansion. The H-I
approach, in combination with lowest order perturbational expansion,
leads to a physically appealing, picture of the Mott transition
including the appearance of a extremely correlated Fermi liquid in its
vicinity, which is complementary to the Gutzwiller-Brinkman-Rice
picture.

It is natural to wonder how well this approach fits as a description
of Mott transitions and the surrounding ECFL in realistic strongly
correlated electron models. In this regard, we emphasize that the
double occupancy conserving Hamiltonian $H_c$ analyzed here is
equivalent to the Hubbard model with bond charge interactions at the
symmetry point $X=-t$ -- which has been argued, {\it e.g.} in
\cite{Strack, Ovchinnikov}, to be a quite realistic value. Our H-I or
Roth improved H-I calculations compare quite favourably with known and
expected results for the latter model, and are remarkably consistent
with them in 1 dimension. Indeed, one would expect the picture
presented here to hold in the neighbourhood of the symmetry point. On
the other hand, for pure Hubbard like--systems, {\it i.e.} $X$ close
to $0$, of course this strong coupling picture of the Mott transition
can only be qualitatively correct (when antiferromagnetic order is
suppressed) and that too only near the transition point.  This gives
rise to an open question concerning the possibility of obtaining an
interpolating scheme between these two extreme cases.

In this regard, we propose to use the following variational ground
state for systems exhibiting metal to Mott insulator transitions:
\[
| \Psi \rangle = \exp [-\alpha \hat{T}_{0}]\exp[-\eta \hat{D}]| \Psi_0
\rangle=\exp [-\alpha \hat{T}_{0}]| {\rm GWF} \rangle,
\]
where $\alpha, \eta$ are variational parameters and $|\Psi_0 \rangle$
is an appropriate reference state.  This state is an extension of the
standard Gutzwiller Wave Function (GWF) and should provide an improved
description of the metallic side of the transition. Note that the two
exponential terms commute and together form the exponential of $-
\alpha H_{c}(U=\eta/\alpha)$, which may be viewed as a partial
projection on to the ground state of $H_{c}(U=\eta/\alpha)$.  This
makes a connection with the ECFL properties described in this paper
and allows for dressing of the GWF with precursors of the Hubbard
bands.  It is worth mentioning here that a similar function
(containing hopping only of the lower Hubbard band in $T_0$) has
already been used in \cite{Baeriswyl} and has provided excellent
results in comparison with exact results for 2 electrons on the
Hubbard square.

\section*{Acknowledgements} We thank Dionys Baeriswyl and  Roman
Micnas for many interesting discussions. We also thank Adam Sajna for
useful comments.   This work  was supported by the Polish National Science Centre under grant
 2011/03/B/ST2/01903. 


\end{document}